\documentclass[aps,pre,twocolumn,eqsecnum]{revtex4}    
\usepackage{amsmath}    % need for subequations
\usepackage{amssymb}
\usepackage{epsfig}

\begin{document}

\title{Non-equilibrium thermodynamics in sheared hard-sphere materials}

\author{Charles K. C. Lieou}
\author{J. S. Langer}
\affiliation{Department of Physics, University of California, Santa Barbara, CA 93106, USA}
\date{\today}

\begin{abstract}
We combine the shear-transformation-zone (STZ) theory of amorphous plasticity with Edwards' statistical theory of granular materials to describe shear flow in a disordered system of thermalized hard spheres. The equations of motion for this system are developed within a statistical thermodynamic framework analogous to that which has been used in the analysis of molecular glasses. For hard spheres, the system volume $V$ replaces the internal energy $U$ as a function of entropy $S$ in conventional statistical mechanics. In place of the effective temperature, the compactivity $X = \partial V / \partial S$ characterizes the internal state of disorder. We derive the STZ equations of motion for a granular material accordingly, and predict the strain rate as a function of the ratio of the shear stress to the pressure for different values of a dimensionless, temperature-like variable near a jamming transition. We use a simplified version of our theory to interpret numerical simulations by Haxton, Schmiedeberg and Liu, and in this way are able to obtain useful insights about internal rate factors and relations between jamming and glass transitions.  
\end{abstract}

\maketitle

\section{Introduction}

In this paper, we combine the basic elements of the shear-transformation-zone (STZ) theory of amorphous molecular plasticity \cite{falk_1998,langer_2011} with Edwards' statistical theory of granular materials \cite{edwards_1989a,edwards_1989b,edwards_1989c,edwards_1990a,edwards_1990b} to construct a theory of shear flow in a noncrystalline system of thermalized hard spheres

The shear-transformation-zone (STZ) theory has been successful in accounting for a range of glassy, nonequilibrium phenomena, such as the formation of shear bands \cite{manning_2007a}, stress-strain curves of metallic glasses in constant strain-rate experiments \cite{langer_2008}, and frequency-dependent viscoelastic response functions \cite{langer_2011a,langer_2011b}. Most recently, one of us (JSL) has proposed that the STZ's play the role of dynamic heterogeneities in producing both Stokes-Einstein violations and stretched-exponential relaxation \cite{langer_2011c}. 

The essence of the STZ theory of plasticity is the assumption that irreversible molecular rearrangements occur at isolated flow defects, i.e.~STZ's. In deforming systems, the STZ's are not fixed structural features.  Rather, they are activated fluctuations that appear and disappear in response to thermal or mechanically generated noise.  The plastic strain rate in this theory is proportional to the STZ density, which is determined by a Boltzmann-like factor of the form $\exp\, ( - e_Z / \theta_{\text{eff}})$, where $e_Z$ is the STZ formation energy, and $\theta_{\text{eff}} = \partial U_C / \partial S_C$ is the effective temperature in energy units. Here, $U_C$ is the configurational potential energy and $S_C$ is the configurational entropy. $\theta_{\text{eff}}$ is a measure of the configurational disorder. 

Sheared, dense, granular materials exhibit features, such as dynamic yield stresses and  jamming transitions, that are similar to those of amorphous molecular systems. However, these systems differ fundamentally from each other because the energies of interaction between ideally hard grains are undefined. For a granular material composed of frictionless hard spheres, there is no potential energy of interaction between grains. Concepts such as the effective temperature $\theta_{\text{eff}}$ and the formation energy $e_Z$ cannot be carried over directly from theories of molecular plasticity. Nevertheless, granular materials do carry entropy, and granular rearrangements under shear must be governed by the second law of thermodynamics just as molecular rearrangements are in an elastoplastic solid. How, then, are we to formulate an STZ theory for granular materials composed of perfectly hard grains?

Edwards and coworkers \cite{edwards_1989a,edwards_1989b,edwards_1989c,edwards_1990a,edwards_1990b} have argued that the total volume $V$ occupied by a granular system plays the role of the configurational energy $U_C$ in the statistical theory of granular materials. In general, like $U_C$, $V$ is a function of the configurational entropy $S_C$ plus other internal state variables to be specified below. The compactivity,
\begin{equation}
\label{eq:compactivity_def}
 X = \dfrac{\partial V}{\partial S_C}
\end{equation}
is the analog of the effective temperature. Our main purpose here is to extend this analogy to nonequilibrium situations.

This paper is structured as follows. We start in Sec.~II by writing out the first and second laws of thermodynamics for a system of hard spheres in contact with a thermal reservoir and driven by external forces. Then, in Secs.~III - V, we derive equations of motion for the STZ's and for the compactivity $X$. In Sec.~VI, we use our theory to interpret the results of molecular dynamics simulations of bidisperse hard spheres by Haxton, Schmiedeberg and Liu \cite{liu_2011,haxton_2012}, hereafter referred to as HSL. An especially important part of the HSL results is Haxton's use of a compressibility identity to measure an effective temperature proportional to $X$  \cite{haxton_2012}, and thereby provide information regarding internal rate factors and relations between jamming and glass transitions. We conclude in Sec.~VII with remarks about the implications of these results.

\section{First and second laws of thermodynamics}\label{sec:2}

Consider a noncrystalline system of hard grains whose kinetic temperature $\theta$ is fixed by contact with a thermal reservoir. For simplicity, assume that this reservoir has a large enough heat capacity that its temperature does not change when energy flows between it and the granular subsystem. Let $U_T$ denote the total energy of this system, including the granular kinetic energies and the energy stored in the reservoir.  If the grains interact only via contact forces, they have no configurational potential energy and, therefore, no such energy is included in $U_T$.  

Suppose that this system is driven in simple (not pure) shear by a shear stress $s$ and a pressure $p$. The first law of thermodynamics for this system is
\begin{eqnarray}
\label{eq:first_law}
 \nonumber \dot{U}_T &=&  V s\, \dot{\gamma} - p\, \dot{V} \\ &=&  V s\, \dot{\gamma} - p X \dot{S}_C - p \sum_{\alpha} \left( \dfrac{\partial V}{\partial \Lambda_{\alpha}} \right)_{S_C} \dot{\Lambda}_{\alpha},
\end{eqnarray}
where $\dot{\gamma}$ is the shear rate, $S_C$ is the granular configurational entropy introduced above in Eq.~(\ref{eq:compactivity_def}), and the $\Lambda_{\alpha}$ are internal variables that specify the configurational state of the granular subsystem.  

Let $S_T$ denote the entropy of the reservoir plus the (quantitatively negligible) entropy of the kinetic degrees of freedom of the grains.  Then
\begin{equation}
 \dot{U}_T = \theta \dot{S}_T ,
\end{equation}
and 
\begin{equation}\label{eq:S_C}
 p X \dot{S}_C = V s \dot{\gamma} - p \sum_{\alpha} \left( \dfrac{\partial V}{\partial \Lambda_{\alpha}} \right)_{S_C} \dot{\Lambda}_{\alpha} - \theta \dot{S}_T.
\end{equation}
The second law of thermodynamics requires that the total entropy be a non-decreasing function of time:
\begin{equation}\label{eq:second_law}
 \dot{S} = \dot{S}_C + \dot{S}_T \geq 0 .
\end{equation}
Substituting Eq.~\eqref{eq:S_C} for $\dot S_C$ into the second law above, and using the fact that each individually variable term in the resulting inequality must be non-negative \cite{langer_2009b,langer_2009c}, we arrive at the second-law constraints
\begin{eqnarray}
 \label{eq:W} {\cal W} = V s\, \dot{\gamma} - p \sum_{\alpha} \left( \dfrac{\partial V}{\partial \Lambda_{\alpha}} \right)_{S_C} \dot{\Lambda}_{\alpha} \geq 0; \\
 \label{eq:S_T} (p X - \theta) \dot{S}_T \geq 0 .
\end{eqnarray}
The dissipation rate ${\cal W}$, as defined in \cite{langer_2009b,langer_2009c}, is the difference between the rate at which inelastic work is done on the configurational subsystem and the rate at which energy is stored in the internal degrees of freedom. The second constraint implies that $p X - \theta$ and $\dot{S}_T$ must carry the same sign if they are nonzero, so that
\begin{equation}\label{eq:Q}
 \theta \dot{S}_T = - {\cal K} \left( 1 - \dfrac{pX}{\theta} \right)\equiv \,{\cal Q},
\end{equation}
where ${\cal K}$ is a non-negative thermal transport coefficient. It is already clear from this analysis that $p\,X$ plays the role of a temperature. $p\,X$ approaches $\theta$ in an equilibrating system; and a heat flux ${\cal Q}$ flows from the granular subsystem into the reservoir when the two subsystems are not in thermodynamic equilibrium with each other.

\section{STZ equations of motion}\label{sec:3}

The following discussion is almost, but not quite, the same as that which has appeared in the STZ literature.~\cite{langer_2011} We repeat it here because several of the differences are important.

As in the case of molecular systems, we assume that the STZ's are two-state flow defects. We further assume that only a single species of STZ is statistically relevant at any given temperature, pressure, or packing fraction.  In other words, for the steady-state behavior of interest here, we do not invoke the explicit distribution over STZ transition rates that was needed in theories of frequency-dependent viscoelasticity \cite{langer_2011a,langer_2011b} or stretched-exponential relaxation \cite{langer_2011c}.  However, some elements of those theories may be relevant here. 

Following ~\cite{falk_1998,langer_2011}, we suppose for simplicity that STZ's can be classified as ``plus'' and ``minus'' according to their orientations relative to the applied shear stress.  We let $N_+$ and $N_-$ denote the number of STZs in each of the two orientations, and let
\begin{equation}
 \Lambda = \dfrac{N_+ + N_-}{N}; \quad m = \dfrac{N_+ - N_-}{N_+ + N_-}
\end{equation}
denote the density and orientational bias of STZ's, where $N$ is the number of grains.

Let $v_Z$ denote the excess volume per STZ. Then the total volume $V$ is
\begin{eqnarray}
 \nonumber V &=& N \Lambda v_Z + V_1 (S_1) \\ &=& N \Lambda v_Z + V_1 (S_C - S_Z (\Lambda, m) ),
\end{eqnarray}
where $V_1$ and $S_1$ are the volume and entropy of all configurational degrees of freedom of the granular system not associated with STZ's, and $S_Z$ is the entropy associated with the STZ's. Then \cite{langer_2009c}
\begin{equation}
 S_Z (\Lambda, m) = N S_0 (\Lambda) + N \Lambda \psi (m)
\end{equation}
where
\begin{eqnarray}
 \label{eq:S_0} S_0 (\Lambda) &=& - \Lambda \ln \Lambda + \Lambda;\\
 \label{eq:psi}\nonumber \psi(m) &=& \ln 2 - \dfrac{1}{2} (1 + m) \ln (1 + m) \\ & & - \dfrac{1}{2} (1 - m) \ln (1 - m).
\end{eqnarray}

The STZ equation of motion for $N_+$ and $N_-$ is generally written in the form:
\begin{equation}\label{eq:master}
 \tau \dot{N}_{\pm} = {\cal R} (\pm s) N_{\mp} - {\cal R} (\mp s) N_{\pm} + \tilde{\Gamma} \left( \dfrac{1}{2} N^{eq} - N_{\pm} \right) .
\end{equation}
The corresponding strain rate is
\begin{equation}
 \dot{\gamma} = \dfrac{2\,v_0}{\tau V} \left[ {\cal R}(s) N_- - {\cal R}(-s) N_+ \right],
\end{equation}
where, because we are describing simple rather than pure shear, we define the volume of the plastic core of an STZ to be $2\,v_0$.

Each term in Eq.~(\ref{eq:master}) needs some interpretation. On the left-hand side, $\tau$ is a time scale that ordinarily has been chosen to be consistent with the underlying micro-scale dynamics, for example, a molecular vibration period. The analogous choice for the hard-sphere system is the inertial time scale $\tau = \sqrt{m/p\,a}$, where $m$ is the average mass of a sphere, and $a^3$ is its average volume.  This is the choice of time scale made by HSL in reporting their data; thus, we adopt it here.  For relatively small packing fractions $\phi = N\,a^3/V$, it can be shown that the inertial $\tau$ is proportional to the average time between sphere-sphere collisions multiplied by a dimensionless function of $\phi$; thus, the inertial time scale is a natural analog of the molecular vibration period.  At larger values of $\phi$, however, where the system becomes jammed, we expect that most of the spheres are in close contact with each other, and that their rearrangement rates --- just as in molecular glasses --- are determined by collective motions that are much slower than $\tau^{-1}$. 

On the right-hand side of Eq.~(\ref{eq:master}), the first two terms containing ${\cal R}(\pm s)$ are the rates (in units of $\tau^{-1}$) at which the STZ's are making forward and backward transitions.  These rates describe much of the basic physics of this class of systems; they are discussed in more detail throughout this paper. The second two terms are the rates of STZ creation and annihilation. These are fluctuation-activated processes, expressed here in the form of a detailed-balance relation in which $N^{eq}$ is the steady-state, total number of STZ's.  

$\tilde\Gamma/\tau$ is an attempt frequency consisting of additive thermal and mechanical parts: 
\begin{equation}
\tilde{\Gamma} = \rho +\Gamma.
\end{equation}
The quantity $\rho$ is best understood as a dimensionless, thermal noise strength.  Well above the jamming transition, where $\tau$ is an accurate approximation for the elementary time scale, $\rho$ should be of the order of unity.  As we approach the jamming transition, either by decreasing $\theta$ or increasing $p$ or $\phi$, $\rho$ decreases rapidly, becoming unmeasureably small below a transition point.  When $\rho=0$, the system is fully jammed in the sense that configurational rearrangements can occur only in response to sufficiently large driving forces. The rate $\rho/\tau$, multiplied by an activation factor of the form $\exp\,(- {\rm const.}/\theta)$, is our analog of the ``$\alpha$'' relaxation rate $\tau_{\alpha}^{-1}$. A first-principles calculation of $\tau_{\alpha}$ is the central unsolved problem of glass physics.  We do not attempt to solve it here; instead, we deduce values of $\rho$ from the HSL data.  

In analogy to $\rho/\tau$, the quantity $\Gamma/\tau$ is the contribution to the attempt frequency in Eq.~(\ref{eq:master}) due to mechanically generated noise.  Here, we do have a well defined prescription for computing $\Gamma$ from the rate of entropy generation, as shown below in Eq.~(\ref{eq:Gamma}).  

In terms of the intensive variables $\Lambda$ and $m$, the STZ equations of motion become:
\begin{eqnarray}
 \label{eq:Lambda} \tau\, \dot{\Lambda} &=& \tilde{\Gamma} ( \Lambda^{eq} - \Lambda ); \\
 \label{eq:m} \tau\, \dot{m} &=& 2\, {\cal C}(s) ( {\cal T}(s) - m ) - \tilde{\Gamma} m - \tau \dfrac{\dot{\Lambda}}{\Lambda} m; \\
 \label{eq:D_pl} \tau \,\dot{\gamma} &=&2\, \epsilon_0\,\Lambda\, {\cal C}(s) ({\cal T}(s) - m),
\end{eqnarray}
where $\epsilon_0 = N\, v_0 / V$ and $\Lambda^{eq} = N^{eq} / N$. We also define
\begin{equation}
\label{eq:calC}
 {\cal C}(s) = \dfrac{1}{2} \left( {\cal R}(s) + {\cal R}(-s) \right) ;
\end{equation}
and
\begin{equation}
 {\cal T}(s) = \dfrac{{\cal R}(s) - {\cal R}(-s)}{{\cal R}(s) + {\cal R}(-s)} .
\end{equation}

At this point in the development, the second law of thermodynamics provides useful constraints on the ingredients of the preceding equations.  Substituting Eqs.~\eqref{eq:Lambda}, \eqref{eq:m} and \eqref{eq:D_pl} into Eq.~\eqref{eq:W} for the dissipation rate, we find
\begin{eqnarray}\label{eq:W2}
 \nonumber &&\tau\, \dfrac{{\cal W}}{N} = - \tilde{\Gamma}\, p\, X\, \Lambda\, m \dfrac{d \psi}{dm} \\\nonumber && + 2\, \Lambda\, {\cal C}(s)\,\Bigl( {\cal T}(s) - m\Bigr) \left( v_0 s + p X \dfrac{d \psi}{dm} \right) \\ && - p\, \tilde{\Gamma}\,(\Lambda^{eq} - \Lambda)  \left[ v_Z + X \left( \ln \Lambda - \psi (m) + m \,\dfrac{d \psi}{dm} \right) \right] .~~~~~~~~~ 
\end{eqnarray}
The second-law constraint, ${\cal W} \geq 0$, implies that each of the three terms in Eq.~\eqref{eq:W2} must be non-negative. The first term automatically satisfies this requirement because, from Eq.~\eqref{eq:psi}, we have
\begin{equation}\label{eq:dpsidm}
 \dfrac{d \psi}{dm} = - \dfrac{1}{2} \ln \left( \dfrac{1 + m}{1 - m} \right) = - \tanh^{-1} (m)
\end{equation}
so that the product $- m ( d \psi / dm)$ is automatically non-negative. 

An especially important result comes from the second term, for which
\begin{equation}\label{eq:Ts}
 \Bigl( {\cal T}(s) - m \Bigr) \left( v_0 s + p X \dfrac{d \psi}{dm} \right) \geq 0.
\end{equation}
The two factors on the left-hand side must be monotonically increasing functions of $s$ that change sign at the same point for arbitrary values of $m$. According to Eq.~\eqref{eq:dpsidm}, this is possible only if
\begin{equation}
 {\cal T}(s) = \tanh \left( \dfrac{v_0 s}{p X} \right).
\end{equation}

The non-negativity constraint on the third term in Eq.~\eqref{eq:W2} can be written in the form
\begin{equation}
 - \dfrac{\partial F}{\partial \Lambda} (\Lambda^{eq} - \Lambda) \geq 0
\end{equation}
where $F$ is a free energy given by
\begin{equation}
 F (\Lambda, m) = p \left[ v_Z \Lambda + X S_0 (\Lambda) - X \Lambda \left( \psi(m) - m \dfrac{d \psi}{d m} \right) \right].
\end{equation}
$\Lambda^{eq}$ must be the value of $\Lambda$ at which $ \partial F / \partial \Lambda$ changes sign, so that
\begin{equation}\label{eq:Lambda_eq}
 \Lambda^{eq} = \exp \left[ - \dfrac{v_Z}{X} + \psi(m) - m \dfrac{d \psi}{dm} \right] \approx 2\, \exp \left( - \dfrac{v_Z}{X} \right). ~~~~~
\end{equation}
Thus, the STZ density in this non-equilibrium situation is given by a Boltzmann-like expression in which the compactivity plays the role of the temperature.

\section{Quasistationary relations}

At this point in our analysis, we specialize to quasistationary or steady-state situations for which we can set $\dot{\Lambda} = \dot{m} = 0$, implying specifically that $\Lambda = \Lambda^{eq}$. We start by assuming that Pechenik's hypothesis \cite{langer_2009c,pechenik_2003} remains valid for a sheared granular material; that is, that the mechanical noise strength $\Gamma$ is proportional to the heat production per STZ. In steady flow, all of the work done on the system is dissipated as heat; therefore the rate of energy dissipation per unit volume is $\dot{\gamma} s$. To convert this rate into a noise strength with dimensions of inverse time, we multiply by the volume per noise source, i.e.~the volume per STZ, $v / \Lambda^{eq}$, and divide by an energy conveniently written in the form $\epsilon_0 v\, s_0$. Here, $v = V/N$, and $s_0$ is a system-specific parameter with the dimensions of stress.  The resulting expression for $\Gamma$ is
\begin{equation}\label{eq:Gamma}
 \Gamma = \dfrac{s\, \dot{\gamma}\, \tau}{\epsilon_0 s_0 \Lambda^{eq}} = \dfrac{2 s}{s_0} {\cal C}(s) \Bigl( {\cal T}(s) - m \Bigr) .
\end{equation}
With this result, the stationary version of Eq.~\eqref{eq:m} reads
\begin{equation}
 2\, {\cal C}(s) \Bigl( {\cal T}(s) - m \Bigr) \left( 1 - \dfrac{m s}{s_0} \right) - m\, \rho  = 0,
\end{equation}
which is satisfied by $m = m^{eq} (s)$, where
\begin{eqnarray}\label{eq:m_eq}
 \nonumber &&m^{eq} (s) = \dfrac{s_0}{2s} \left[ 1 + \dfrac{s}{s_0} {\cal T}(s) + \dfrac{\rho}{2\, {\cal C}(s)} \right] \\ &&- \dfrac{s_0}{2s} \sqrt{\left[ 1 + \dfrac{s}{s_0} {\cal T}(s) + \dfrac{\rho}{2\, {\cal C}(s)} \right]^2 - 4 \dfrac{s}{s_0} {\cal T}(s)}. ~~~~~
\end{eqnarray}
In particular, when $\rho = 0$, we find
\begin{equation}
 m^{eq} = 
\begin{cases}
{\cal T}(s), & \text{if $(s / s_0)\, {\cal T}(s) < 1$} ; \\
s_0/s, & \text{if $(s / s_0)\, {\cal T}(s) \geq 1$}.
\end{cases}
\end{equation}
Thus an exchange of stability occurs in the same manner as it did in molecular systems, with the low-temperature yield stress being the solution of the equation
\begin{equation}
 s_y {\cal T}(s_y) = s_y \tanh \left( \dfrac{v_0 s_y}{p X}\right) = s_0.
\end{equation}
If the temperature-like quantity $p\,X$ is small in comparison with $v_0 s_0$, then $s_y \approx s_0$.

Finally, the steady-state version of Eq.~\eqref{eq:D_pl} for the strain rate becomes
\begin{equation}\label{eq:q_s}
 q \equiv \tau \dot{\gamma} = 4\,\epsilon_0\,e^{-\,1/\chi}\,{\cal C}(s) \left[\tanh \left( \dfrac{v_0 s}{v_Z\,p\,\chi} \right) - m^{eq} (s) \right] .
\end{equation}
Here, we have introduced the inertial number $q$ as a dimensionless measure of the strain rate.  (E.g. see \cite{jop_2006}.)   We also have introduced the dimensionless compactivity $\chi = X/v_Z$. 

\section{Kinematic equations for the compactivity}

The dimensionless compactivity $\chi$ is the analog of the effective temperature in molecular systems and, as such, is a measure of the system's state of structural disorder. Referring to Eq.~\eqref{eq:Q}, we see that the natural definition of a thermal temperature comparable to $\chi$ is
\begin{equation}
\label{eq:tildetheta}
 \tilde{\theta} \equiv \dfrac{\theta}{p\, v_Z},
\end{equation}
which is almost the same as the dimensionless temperature $\tilde\theta_H = \theta/p\,a^3$ defined by HSL. 

Below the jamming transition, i.e. at low temperatures or high pressures where $\rho = 0$, and where long-lived structural rearrangements can be induced only by externally driven deformation, we expect  there to be a direct relation between the steady-state shear rate and the compactivity, say, $\chi = \hat{\chi}(q)$. When the shear rate is much smaller than any relevant relaxation rate in the system, then dimensional analysis requires that $\hat{\chi}(q)$ be a $q$-independent constant, say $\hat{\chi}_0$. However, when the shear rate becomes comparable to internal relaxation rates, i.e., when the system is being ``stirred'' rapidly on its intrinsic time scales, its disorder increases, and $\hat{\chi}(q)$ becomes an increasing function of $q$. The simplest way to describe this situation is to write $\hat\chi$ as a function of the ratio $q/q_0$, where $q_0/\tau$ is an internal rate relevant to shear relaxation.  One obvious possibility is that $q_0$ is proportional to ${\cal R}(0)$, which is a measure of the rate at which STZ's spontaneously undergo shear transitions between their two states. This rate becomes slow in jammed systems, but it generally remains nonzero even when $\rho = 0$. 

In recent publications \cite{manning_2007b,langer-egami-2012}, it has been assumed that the large-$q$ relation between $q/q_0$ and $\hat{\chi}$ has a Arrhenius form, $q/q_0 \sim \exp\, (- A / \hat{\chi})$ for $\hat\chi \gg \hat\chi_0$. Both $\hat{\chi}$ and $A$ are volumes measured in units of $v_Z$. If $v_Z$ is the only volume in the system relevant to configurational rearrangements, then we expect that $A$ is of the order of unity.  A relation of this form was discovered in numerical simulations by Haxton and Liu~\cite{haxton_2007}. 

To interpolate between the small-$q$ and large-$q$ behaviors of $\hat\chi(q)$, it was proposed in \cite{manning_2007b} that $\hat\chi(q)$ be written phenomenologically in the form of a Vogel-Fulcher-Tamann (VFT) expression for a ``viscosity'' $q^{-1}$ as a function of the temperature $\hat{\chi}$:
\begin{equation}\label{eq:q_xhat}
 \dfrac{1}{q} = \dfrac{1}{q_0} \exp \left[ \dfrac{A}{\hat{\chi}} + \alpha_{eff} ( \hat{\chi} ) \right],
\end{equation}
with,
\begin{equation}\label{eq:alpha_eff3} 
 \alpha_{\text{eff}} ( \hat{\chi} ) = \left( \dfrac{\hat{\chi}_1}{\hat{\chi} - \hat{\chi}_0} \right) \exp \left( -b\, \dfrac{\hat{\chi} - \hat{\chi}_0}{\hat{\chi}_A - \hat{\chi}_0} \right) .
\end{equation}
Thus, $\hat{\chi} \rightarrow \hat{\chi}_0$ in the limit of small strain rates, and $\hat{\chi} \rightarrow \infty$ as $q \rightarrow q_0$. The exponential cutoff in Eq.~\eqref{eq:alpha_eff3} is needed in order that the VFT divergence at small $\hat{\chi}$ transforms smoothly to the Arrhenius law at large $\hat{\chi}$. Previous calculations have used $b = 3$.

The equation of motion for $\chi$ itself is a statement of the first law of thermodynamics; it describes entropy flow through the slow, configurational degress of freedom into the fast thermal motions of the grains. Near steady state, it has the form
\begin{equation}\label{eq:chi_motion}
 \dot{\chi} \propto e^{-1/\chi} \left[ \Gamma \Bigl( \hat{\chi}(q) - \chi \Bigr) + \kappa\, \rho\, \Bigl(\tilde{\theta} - \chi\Bigr) \right].
\end{equation}
The first term in the square brackets on the right-hand side is the rate at which $\chi$ is driven towards $\hat{\chi}(q)$ by the mechanical noise strength $\Gamma$. The second term, proportional to $\rho$, is the rate at which thermal fluctuations drive $\chi$ toward the scaled ambient temperature $\tilde{\theta}$. $\kappa$ is a dimensionless parameter of the order of unity. The competition between the two terms in Eq.~\eqref{eq:chi_motion} determines the value of $\chi$; it is close to $\hat{\chi}(q)$ for large $\Gamma$, and close to $\tilde{\theta}$ when the system is driven slowly so that $\Gamma$ is small.

There is one complication that must be dealt with at this point. Equation~\eqref{eq:chi_motion}, as written, implies that the steady-state $\chi$ must lie in the interval between $\hat{\chi}(q)$ and $\tilde{\theta}$. If we assume that $\hat{\chi}(q) \approx \hat{\chi}_0$ is a constant for small enough $q$, then a system initially prepared with $\tilde{\theta} > \hat{\chi}_0$ would be ``cooled'' to $\chi < \tilde{\theta}$ when driven at a small strain rate. This behavior seems implausible; so far as we know, it is not seen in simulations, e.g. \cite{haxton_2012}, in which $\chi$ is measured directly. In \cite{manning_2007b}, this problem was corrected by setting $\hat{\chi}_0 = \tilde{\theta}$ when $\tilde{\theta}$ exceeds $\chi_0$, and by rescaling $\hat{\chi}_1$ and $\hat{\chi}_A$ accordingly. Specifically,
\begin{equation}
\label{chi0}
\hat{\chi}_0 = 
 \begin{cases} 
 \chi_0, & \text{if $\tilde{\theta} < \chi_0$}; \\
 \tilde{\theta}, & \text{if $\tilde{\theta} > \chi_0$},
 \end{cases}
\end{equation}
and
\begin{eqnarray}
\label{chi1-chiA}
\nonumber
\hat{\chi}_1 &=&
\begin{cases}
\chi_1 , & \text{if $\tilde{\theta} < \chi_0$}; \\
\tilde{\theta} \chi_1 / \chi_0 , & \text{if $\tilde{\theta} > \chi_0$}.
\end{cases}
\\ 
\hat{\chi}_A &=&
\begin{cases}
\chi_A , & \text{if $\tilde{\theta} < \chi_0$}; \\
\tilde{\theta} \chi_A / \chi_0 , & \text{if $\tilde{\theta} > \chi_0$}.
\end{cases}
\end{eqnarray}
This guarantees that $\hat{\chi}(q) > \tilde{\theta}$ at all times.

In summary, we use Eq.~\eqref{eq:chi_motion} in the form
\begin{equation}\label{eq:chi_steady}
 \chi = \dfrac{\Gamma \hat{\chi}(q) + \kappa\, \rho\, \tilde{\theta}}{\Gamma + \kappa\, \rho},
\end{equation}
along with Eq.~\eqref{eq:q_xhat}, to determine one relation between $\chi$, $q$ and $s$. We then use Eq.~\eqref{eq:q_s} to compute both $q$ and $\chi$ as functions of $s$.

\section{Analysis of the HSL  simulations}

In their molecular dynamics simulations, HSL studied mixtures of 4096 hard spheres, half each of size $a$ and $1.4\, a$, in contact with a thermal reservoir at fixed temperature $\theta = 1$, undergoing steady-state, simple shear driven by a shear stress $s$ and a pressure $p$. They reported the strain rates as functions of $s / p$ for a wide range of dimensionless temperatures 
\begin{equation}
\label{thetaH}
\tilde\theta_H = {\theta \over p\, a^3} = {v_Z\over a^3}\,\tilde\theta
\end{equation}
and packing fractions $\phi$. In a separate paper~\cite{haxton_2012}, Haxton reported measurements of the  compressibility temperature $\theta_{\text{comp}}$ of this system using the relation
\begin{equation}\label{eq:comptemp_def}
 (N/V)\, K\, \theta_{\text{comp}} = S( k \rightarrow 0)
\end{equation}
where $K$ is the isothermal compressibility and $S(k)$ is the structure factor as a function of the wavenumber $k$.  He argued that $\theta_{\text{comp}}$ is the thermodynamic temperature of the configurational degrees of freedom for this system, and therefore ought to be the configurational effective temperature in nonequilibrium situations where $\theta_{\text{comp}} \ne \theta$.  Accordingly, in analogy to Eq.~(\ref{eq:tildetheta}), we assume that 
\begin{equation}
\chi = {\theta_{\text{comp}}\over p\,v_Z}.
\end{equation} 

%%%%%%%%%%%%% FIGURE 1 %%%%%%%%%%%%
\begin{figure}[here]
\centering \epsfig{width=.45\textwidth,file=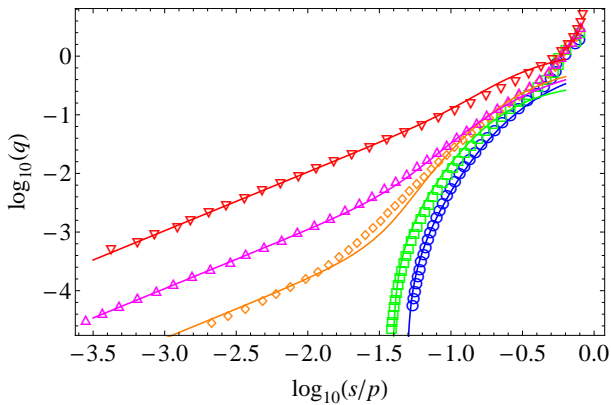} \caption{(Color online) Log-log plots of simulated (data points) and theoretical (solid curves) values of the dimensionless strain rates $q = \tau \dot{\gamma}$ as functions of $s/p$, for dimensionless temperatures $\tilde\theta_H = 0.03$ (blue circles), 0.05 (green squares), 0.08 (orange diamonds), 0.10 (magenta triangles) and 0.25 (red down triangles), shown in that order from bottom to top.} 
\end{figure}
%%%%%%%%%%%%%%%%%%%%%%%%%%%%%%%%%%%%%  

A selection of five of the thirteen HSL data sets is shown in Figures 1 - 3, along with our corresponding theoretical results.  Figures 1 and 2 are log-log plots, respectively, of $q$ and the dimensionless, viscosity-like ratio $s/p\,q$ as functions of $s/p$, for the values of $\tilde\theta_H$ indicated in the figure caption.  In Fig. 3 we show graphs of  
\begin{equation}
\label{chiH}
\chi_H \equiv  {\theta_{\text{comp}}\over p\,a^3} = {v_Z\over a^3}\,\chi
\end{equation}
as functions of $\log_{10}(q)$, again for five selected values of  $\tilde\theta_H$. 

The following analysis is closely parallel to that which has been used for interpreting molecular dynamics simulations of shear flow in a metallic glass \cite{langer-egami-2012}.  In \cite{langer-egami-2012}, the system undergoes an ordinary thermal glass transition, but there were no direct measurements of the thermodynamic effective temperature with which to compare the theory. As will be seen here, the measurements of $\chi_H$ shown in Fig.3 provide crucially important information. 

We find that we can fit the HSL data with a highly simplified version of the STZ theory outlined above in Secs. III - V. Our first major assumption is that the quantities $\rho$, $s_0$, ${\cal R}(0)\equiv R_0$ and $q_0$ are functions only of the scaled temperature $\tilde\theta$ defined in Eq.~(\ref{eq:tildetheta}). 

Second, we assume that the symmetric function ${\cal C}(s)$ defined in Eq.~(\ref{eq:calC}) is simply a constant, i.e. ${\cal C}(s) \approx{\cal C}(0) = R_0(\tilde\theta)$.  We further assume that the inertial rate $\tau^{-1}$ is a sufficiently accurate estimate of the attempt frequency for high-temperature, activated processes that we can let $R_0(\tilde\theta) \rightarrow 1$ and $\rho(\tilde\theta) \rightarrow 1$ in the limit of large $\tilde\theta$. We do not attempt to predict the temperature dependence of either of these fundamentally glassy rate factors at low temperatures; rather, they are regarded here as measurable quantities. 

Our third physically significant assumption is that the ratio $a^3/v_Z$ is a constant, independent of $\tilde\theta$.  In other words, we assume that the activation volume relevant to STZ creation is some fixed multiple of the volume of a grain, independent of whether the system is above or below a jamming transition, independent of whether the transition rate $R_0(\tilde\theta)$ is fast or slow, and even independent of whether the STZ's are compact or extended objects under such different circumstances.  Our idea is that $a^3$ is a measure of the free volume that must be created in order for grains to exchange places with each other, and that it does not matter whether that volume is created quickly in a small, loose cluster of grains, or whether it is created more slowly via rare fluctuations in a larger region. For similar reasons, we assume that $2\,v_0$, the volume of the plastic core of an STZ, is also a fixed multiple of $a^3$.

%%%%%%%%%%%%% FIGURE 2 %%%%%%%%%%%%
\begin{figure}[here]
\centering \epsfig{width=.45\textwidth,file=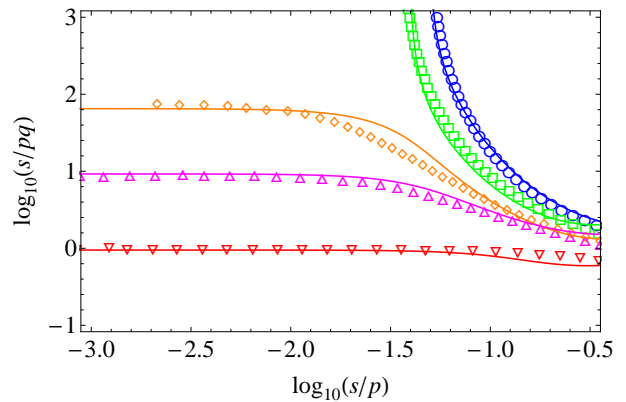} \caption{(Color online) Log-log plots of simulated (data points) and theoretical (solid curves) values of the dimensionless viscosities $s/p\,q$  as functions of $s/p$, for dimensionless temperatures $\tilde\theta_H = 0.03$ (blue circles), 0.05 (green squares), 0.08 (orange diamonds), 0.10 (magenta triangles) and 0.25 (red down triangles), shown in that order from top to bottom.} 
\end{figure}
%%%%%%%%%%%%%%%%%%%%%%%%%%%%%%%%%%%%%  

Finally, the dimensionless numbers $\epsilon_0$ and $\kappa$, which first appeared in Eqs.~\eqref{eq:D_pl} and \eqref{eq:chi_motion}, should both be of the order of unity. In the absence of better information, we set $\epsilon_0 = \kappa = 1$.

The ratio $v_Z/a^3$ is especially important for our purposes, because it is the factor needed to convert from our temperature variables $\tilde\theta$ and $\chi$ to the HSL variables $\tilde\theta_H$ and $\chi_H$, as shown in Eqs.~(\ref{thetaH}) and (\ref{chiH}).  We can determine this ratio, as well as the value of $v_0/v_Z$, by looking at the HSL data in the limit of high temperatures and small stresses, where --- as seen in Figs. 1 and 2 --- the system exhibits simple linear viscosity. In this limit, we know from Eq.~\eqref{eq:chi_steady} that $\chi \approx \tilde{\theta}$.  We also know by taking the small-stress limit of Eqs.~(\ref{eq:m_eq}) and (\ref{eq:q_s}), and setting $\rho = R_0 =1$, that 
\begin{equation}
\label{eq:viscosity}
 \dfrac{s}{p\,q} \approx \dfrac{3\,\tilde{\theta}\,v_Z}{4\, v_0}\, e^{1/\tilde\theta}. 
\end{equation}
Evaluating the left-hand side of this equation directly from the HSL data at temperatures $\tilde\theta_H = 0.25$ and $0.30$, we find that $v_Z/a^3 = \tilde\theta_H/\tilde\theta \cong 0.53$ and $v_0/a^3 \cong 1.65$.  Thus, the activation volume $v_Z$ is somewhat smaller than the average grain size, and the plastic core of an STZ is somewhat bigger. Both are plausible results; but we emphasize that both depend on our assumption about the high-temperature limits of $R_0$ and $\rho$.

%%%%%%%%%%%%% FIGURE 3 %%%%%%%%%%%%
\begin{figure}[here]
\centering \epsfig{width=.45\textwidth,file=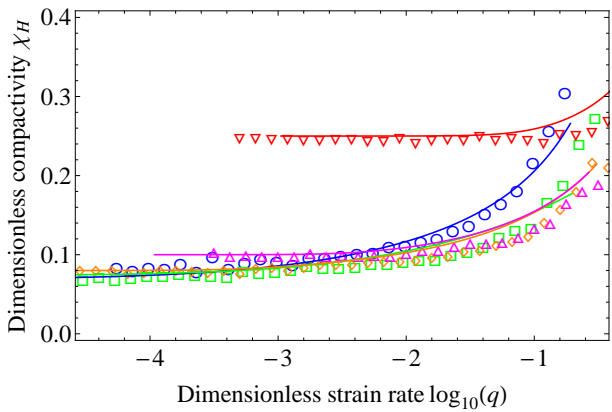} \caption{(Color online) Log-linear plots of simulated (data points) and theoretical (solid curves) values of the effective temperature $\chi_H$ as functions of the dimensionless strain rate $q$, for dimensionless temperatures $\tilde\theta_H = 0.03$ (blue circles), 0.05 (green squares), 0.08 (orange diamonds), 0.10 (magenta triangles) and 0.25 (red down triangles). } 
\end{figure}
%%%%%%%%%%%%%%%%%%%%%%%%%%%%%%%%%%%%%  

The next step in our analysis is to look at low temperatures, in the jammed region, where $\rho = 0$ and the data seem to indicate well defined yield stresses. In this limit, $\chi = \hat\chi(q)$, and Eqs.~(\ref{eq:m_eq}) and (\ref{eq:q_s}) become 
\begin{equation}
\label{eq:qlow}
 q = 4\,\, R_0(\tilde\theta)\, e^{-1 / \hat\chi(q)} \left(1 - \dfrac{s_0(\tilde\theta)}{s} \right).
\end{equation}
This equation allows us to evaluate the function $s_0(\tilde\theta)$ directly from the observed yield stresses.  More interestingly, Eq.~(\ref{eq:q_xhat}) tells us that $\hat\chi$ is a universal function of the ratio $q/q_0(\tilde\theta)$ so long as the parameters $A$, $\hat\chi_0$, etc. appearing there and in Eq.~(\ref{eq:alpha_eff3}) are independent of $\tilde\theta$, which we assume to be true.  Therefore, in this low-temperature regime, graphs of $\chi(q)$ should collapse to a single curve when shifted along the $\log_{10}(q)$ axis by amounts $\log_{10}(q_0)$.  The lower graphs shown in Fig.~3 do behave in roughly this way.  Thus, by fitting these curves and using Eq.~(\ref{eq:qlow}) to fit the low-temperature graphs in Figs.~1 and 2, we can determine the temperature-independent quantities $A$, $\chi_0$, $\chi_1$, and $\chi_A$, defined in Eqs.~(\ref{chi0}) and (\ref{chi1-chiA}); and we can determine the low-temperature values of the functions $R_0(\tilde\theta)$ and $q_0(\tilde\theta)$.  Specifically, we find that $A \cong 1.09 $, $\chi_0 \cong 0.12$, $\chi_1 \cong 0.038$, and $\chi_A \cong 0.47$.  

Having evaluated the temperature-independent parameters, it is a straghtforward exercise to complete this data analysis using the full STZ equations.  We need to fit the data in the intermediate regions between fully jammed ($\rho = 0$) and fully unjammed ($\rho = 1$) situations.  We also need to look in more detail at situations in which the driving forces $s/p$ and the resulting strain rates $q$ become large.  In the latter cases, we know that the STZ theory must fail if the density of STZ's becomes too large and the system begins to behave more like a liquid than a solid.  The HSL data shown in Fig.~1 do show signs of softening at the highest strain rates, and we make no attempt to fit such behavior.  (In \cite{langer-egami-2012}, it was found that this behavior at large stresses and low temperatures could be accounted for in part  by addding a Bagnold-like stress dependence to the STZ transition rate.)  On the other hand, the high-temperature data indicate nonlinear response at the larger values of $q$.  This nonlinearity is sensitive to $s_0(\tilde\theta)$; thus we are able to evaluate that function outside the jammed region.  
  
%%%%%%%%%%%%% FIGURE 4 %%%%%%%%%%%%
\begin{figure}[here]
\centering \epsfig{width=.45\textwidth,file=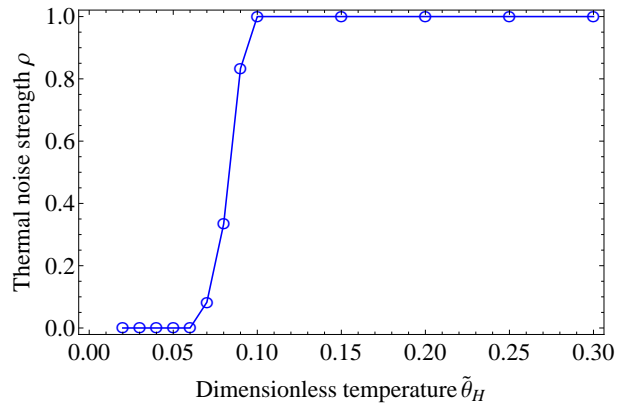} \caption{(Color online) Thermal noise strength $\rho$ as a function of the dimensionless temperature $\tilde\theta_H$.} 
\end{figure}
%%%%%%%%%%%%%%%%%%%%%%%%%%%%%%%%%%%%%  

The function $\rho(\tilde\theta)$, shown in Fig.4 as a function of $\tilde\theta_H$, indicates that the hard-sphere system undergoes a jamming transition in the region $0.05 < \tilde\theta_H < 0.10$.  Below those values of $\tilde\theta_H$, the rate factor $\rho$ becomes unmeasureably small; but we have no way to determine whether or not it actually vanishes.  The internal transition rate $R_0(\tilde\theta)$, shown as a function of $\tilde\theta_H$ in Fig.~5, undergoes a qualitatively similar transition at smaller values of $\tilde\theta_H$, but remains nonzero down to the lowest temperatures for which data is available. Consistent with our discussion in Sec.~V, the independently determined low-temperature values of $q_0(\tilde\theta)$  are roughly proportional to $4\,R_0(\tilde\theta)$, as shown in Fig.~5. This factor $4$ might be the same factor that appears in Eq.~(\ref{eq:qlow}); but we emphasize that Eq.~(\ref{eq:q_xhat}), in which $q_0$ was defined operationally, is no more than a phenomenological interpolation formula that should not be interpreted too literally. At higher temperatures, where $\chi \approx \tilde\theta$, $q_0(\tilde\theta)$ plays no role in the theory and therefore, in principle, cannot be evaluated.   The stress scale $s_0$ shown in Fig.~6 decreases as a function of $\tilde\theta_H$ within the jammed region, and then becomes a constant at higher temperatures.  This is qualitatively the same behavior that was observed in \cite{manning_2007b} and \cite{langer-egami-2012}. 

%%%%%%%%%%%%% FIGURE 5 %%%%%%%%%%%%
\begin{figure}[here]
\centering \epsfig{width=.45\textwidth,file=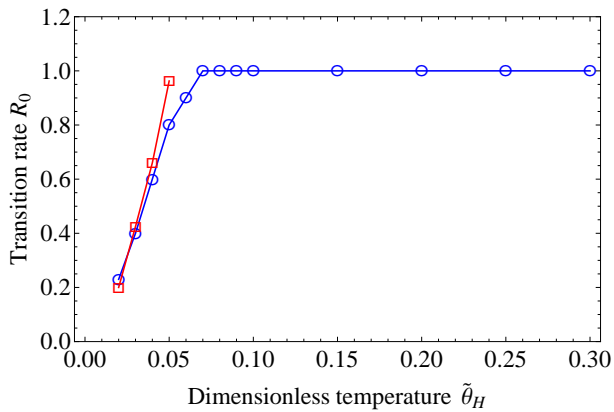} \caption{(Color online) Dimensionless STZ transition rate $R_0$ (blue circles) as a function of the dimensionless temperature $\tilde\theta_H$. The red squares indicate independently determined values of $q_0/4$ at the four lowest values of $\tilde\theta_H$.  Our rough estimates indicate that $q_0$ levels off at about $4.5\,R_0$ above $\tilde\theta_H \approx 0.06$, but is intrinsically not measureable at appreciably higher temperatures.} 
\end{figure}
%%%%%%%%%%%%%%%%%%%%%%%%%%%%%%%%%%%%% 

As shown in Figs.~1 and 2, the resulting fits to the data for strain rates as functions of stress are quite good everywhere except at the largest values of $q$ where, as noted earlier, we expect the STZ theory to fail. The graphs of $\chi_H$ versus $q$ in Fig.~3 are harder to decipher, in part because the data are noisier and, in part, because the small-$q$ behaviors happen to lie on top of one another.  The two curves that are in the jammed region where $\rho = 0$, with $\tilde\theta_H = 0.03$ and $0.05$, are slowly (logarithmically) approaching $\chi_H \cong 0.53\,\chi_0 \cong 0.064$; but they are still above that value at the smallest $q$'s. The next two curves, for $\tilde\theta_H = 0.08$ and $0.10$, already have reached their asymptotic values at $\chi_H = \tilde\theta_H$, because $\Gamma \ll \rho$ in Eq.~(\ref{eq:chi_steady}).  As a result, the four lower curves in the Figure appear to be indistinguishable but, in fact, are consistent with the theory and reveal the importance of Haxton's effective-temperature data for understanding the competing time scales in this system.  The single, distinguishably high-temperature curve in Fig.~3 is nearly flat at $\chi_H \cong \tilde\theta_H \cong 0.25$.  The fact that this curve is crossed by the lowest-temperature curve with $\tilde\theta_H \cong 0.03$ illustrates the sharp difference in the dynamics between the jammed and unjammed regions.

\section{Concluding Remarks}

The thermalized hard-sphere model studied numerically by HSL is an ideal laboratory for studying glass-like behavior near a jamming transition.  It is sharply defined, with no ambiguities about the role of interaction energies or about the origin of the underlying time scale.  As shown in Sec.~II, it lends itself easily to a  nonequilibrium thermodynamic analysis that systematically constrains the STZ equations of motion and relates the Edwards compactivity $X$ to the thermodynamic effective temperature $\chi$.  

The HSL data is especially revealing because of Haxton's direct measurements of the effective temperature \cite{haxton_2012}, without which we could not have sorted out the temperature dependences of the different rate factors that appear in the STZ theory.  For example, there was no such information about $\chi$ in the metallic glass simulations used by Egami and JSL \cite{langer-egami-2012}, and therefore no reason to expect that the strain-rate parameter $q_0$ defined in Sec.~V was a strongly varying function of temperature.  As a result, it was concluded there that the discrepancies between the theoretical and simulated viscosities at small strain rates and temperatures near the glass transition were due to a failure of the single-species STZ theory.  Here we see no such strong discrepancy, presumably because we are taking more accurate account of the $\tilde\theta$ dependence of the STZ transition rates in both $R_0(\tilde\theta)$ and $q_0(\tilde\theta)$.  This speculation, however, leads to yet more interesting questions.

%%%%%%%%%%%%% FIGURE 6 %%%%%%%%%%%%
\begin{figure}[here]
\centering \epsfig{width=.45\textwidth,file=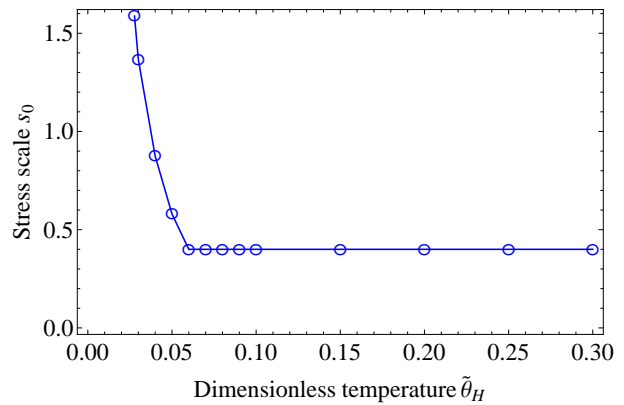} \caption{(Color online) Stress scale $s_0$ in units of $\theta/a^3$ as a function of the dimensionless temperature $\tilde\theta_H$.} 
\end{figure}
%%%%%%%%%%%%%%%%%%%%%%%%%%%%%%%%%%%%%  

The most important of these questions, in our opinion, pertain to the relations between the present results and the postulate in \cite{langer_2011c} that the STZ's are the dynamic heterogeneities frequently invoked as explanations for a wide range of glassy phenomena. In \cite{langer_2011a,langer_2011b}, Bouchbinder and JSL showed that both basic statistical principles and the results of frequency-dependent viscoelastic measurements implied the existence of a broad spectrum of STZ transition rates in equilibrated systems near their glass temperatures. In \cite{langer_2011c}, it was argued that the ``slow'' STZ's near the bottom of this multi-species STZ spectrum are responsible for anomalously large viscosities consistent with observed Stokes-Einstein violations; and that the extended spectrum naturally produces stretched-exponential relaxation in a variety of experimental situations. 

Here, we have used only a single-species STZ theory; but we can interpret our values of $R_0(\tilde\theta)$ as weighted averages over the full spectrum in the multi-species theory.  A difficulty with this interpretation is that, according to the earlier statistical analysis, the spectrum should shift to higher transition rates as the system becomes noisier at larger driving stresses; but, so far, we have assumed that $R_0(\tilde\theta)$ is independent of the stress.  Nevertheless, the present results suggest that the statistically effective STZ's become markedly slower as $\tilde\theta$ decreases through the jamming transition. Recently, Durian and coworkers \cite{durian_2007,durian_2011}, in experiments using air-fluidized grains, have found that their dynamical heterogeneities become slower as the jamming transition is approached.  They also have found that these slower heterogeneities involve motions of larger numbers of grains, a size effect that seems especially intriguing.  It was argued in \cite{langer_2011c} that the length of the diffusion jumps associated with slow STZ's must grow with decreasing temperature in order to account for the apparently normal, Fickian diffusion observed in those conditions.  It will be important to find out whether these are related phenomena.

\section*{Acknowledgments}

We thank Thomas Haxton, Michael Schmiedeberg, and Andrea Liu for generously sharing their simulation data with us. We also thank Ahmed Elbanna and Jean Carlson for instructive discussions. CKCL was supported by an Office of Naval Research MURI grant N000140810747, NSF grant DMR0606092, and the NSF/USGS Southern California Earthquake Center, funded by NSF Cooperative Agreement EAR?0529922 and USGS Cooperative Agreement 07HQAG0008, and the David and Lucile Packard Foundation. JSL was supported in part by the Division of Materials Science and Engineering, Office of Basic Energy Sciences, Department of Energy, DE-AC05-00OR-22725, through a subcontract from Oak Ridge National Laboratory.

\end{document}